\documentclass[10pt,aps,pra,twocolumn,superscriptaddress,floatfix,nofootinbib]{revtex4-2}
\makeatletter
\def\@bibdataout@aps{%
 \immediate\write\@bibdataout{%
  @CONTROL{%
   apsrev41Control,author="08",editor="1",pages="0",title="0",year="1",eprint="1"%
  }%
 }%
 \if@filesw
  \immediate\write\@auxout{\string\citation{apsrev41Control}}%
 \fi
}%
\makeatother 

\usepackage{lipsum}
\usepackage{amsmath}
\usepackage{amsfonts, amssymb, bbm, braket, accents}
\usepackage{graphicx}   
\usepackage[usenames,dvipsnames]{color}
\usepackage[makeroom]{cancel}
\usepackage{soul} 
\usepackage{tikz} 
\usetikzlibrary{shapes.arrows} 
\usetikzlibrary{shapes.geometric}
\usetikzlibrary{calc} 
\usepackage[makeroom]{cancel}
\usepackage{verbatim}
\usepackage{pifont}
\usepackage[mathscr]{euscript}

\newcommand{\brak}[1]{\left\langle #1\right\rangle}
\newcommand{\nn}{\nonumber}
\newcommand{\dg}{^\dagger}
\newcommand{\smallfrac}[2]{\mbox{$\frac{#1}{#2}$}}
\newcommand{\half}{\smallfrac{1}{2}}
\newcommand{\ketbra}[2]{\left\vert{#1}\right\rangle\!\left\langle{#2}\right\vert}

\newcommand{\id}{\mathbbm{1}}
\newcommand{\Id}{\openone}
\newcommand{\ip}[2]{\left\langle{#1}|{#2}\right\rangle}
\newcommand{\expt}[1]{\langle {#1} \rangle}

\newcommand{\op}[2]{\left |{#1}\right\rangle\! \!\left \langle {#2}\right |}
\newcommand{\defeq}{\mathrel{:=}}
\newcommand{\comm}[2]{\left[#1,#2\right]}
\newcommand{\modulus}[1]{\left\vert#1\right\vert}

\newcommand{\p}[0]{p}
\newcommand{\x}[0]{x}
\newcommand{\fin}{f_{\rm in}}
\newcommand{\xbin}{x_{b,{\rm in}}}
\newcommand{\pbin}{p_{b,{\rm in}}}
\newcommand{\ain}{a_{\rm in}}
\newcommand{\aout}{a_{\rm out}}
\newcommand{\bin}{b_{\rm in}}
\newcommand{\bout}{b_{\rm out}}
\newcommand{\xin}{x_{\rm in}}
\newcommand{\xout}{x_{\rm out}}
\newcommand{\pin}{p_{\rm in}}
\newcommand{\pout}{p_{\rm out}}

\usepackage[breaklinks=true]{hyperref}
\hypersetup{
  colorlinks   = true, 
  urlcolor     = blue, 
  linkcolor    = blue, 
  citecolor   = red 
}
\usepackage{amsthm}

\usepackage[capitalise]{cleveref} 
\crefformat{equation}{Eq.~(#2#1#3)} 
\crefformat{section}{Sec.~#2#1#3} 
\Crefformat{equation}{Equation~(#2#1#3)}
\crefformat{figure}{Fig.~#2#1#3}
\crefrangeformat{equation}{Eqs.~#3(#1)#4--#5(#2)#6}
\Crefformat{section}{Section~#2#1#3}

\begin{document}
\title{Quantum limits on noise for a class of nonlinear amplifiers}

\author{Jeffrey M. Epstein}
\affiliation{Berkeley Center for Quantum Information and Computation, Berkeley, CA 94720, USA}
\affiliation{Department of Physics, University of California, Berkeley, CA 94720, USA}

\author{K. Birgitta Whaley}
\affiliation{Berkeley Center for Quantum Information and Computation, Berkeley, CA 94720, USA}
\affiliation{Department of Chemistry, University of California, Berkeley, CA 94720, USA}

\author{Joshua Combes}
\affiliation{Rigetti Computing, 775 Heinz Avenue, Berkeley, CA 94710, USA}
\affiliation{Department of Electrical, Computer and Energy Engineering, University of Colorado Boulder, Boulder, Colorado 80309, USA}

\date{\today}

\begin{abstract}
Nonlinear amplifiers such as the transistor are ubiquitous in classical technology,
but their quantum analogues are not well understood. We introduce a class of nonlinear amplifiers that amplify any normal operator and  add only a half-quantum of vacuum noise at the output. In the large-gain limit, when used in conjunction with a noisy linear detectors, these amplifiers implement ideal measurements of the normal operator.
\end{abstract}

\maketitle

\section{Introduction}\label{sec:intro}
The fundamental physical limit on signal amplification, linear or otherwise, arises from quantum theory, in particular from the requirement of unitarity.
The noise properties of quantum linear amplifiers, devices for which the output signal is linearly related to the input signal, are well understood~\cite{HausMull62,Cave82,ClerDevoGirv10,CaveCombJian12, Aumentado2020}. 
Linear amplifiers that come close to adding the minimum amount of noise i.e., one half units of number quanta, are called {\em quantum-limited}. Such quantum-limited amplifiers have become the measurement workhorse for quantized microwave circuits and mechanical oscillators~\cite{YurkeKami88,*NEC_RIKEN08,*JILA09,*Yale10,*UCB_MIT15, *Chalmers15}. They also show promise for signal transduction~\cite{LauClerk2019,Zeuthen_2020} and
for the advancement of fundamental science, e.g., in the axion dark-matter experiment \cite{KiniClar11}. Although many of these amplifiers are fundamentally nonlinear~\cite{BogdanFederov2015}, they are typically operated in linear regimes where residual nonlinearities may be treated as imperfections and so they can be treated with a linear theory.

While amplifiers that operate in a non-linear regime are not common in quantum technologies, genuinely nonlinear amplifiers are widely used in classical electronics, e.g., for rectification, switching, and logic.
A genuine nonlinear amplifier is a device in which the output is not linear in some property of the input signal such as signal amplitude. 
In some situations, it is known that such nonlinear amplifiers can add less noise than linear amplifiers \cite{Chen69}.
This suggests that nonlinear amplifiers may be a useful quantum technology,  if it is possible to formulate a consistent theory that takes into account the quantum nature of the noise. 
In addition, a systematic theory of quantum nonlinear amplifiers could help elucidate quantum limits on switching and classical logic \cite{Mabu11a,*Mabu11b}.

Despite this promise, the analysis of nonlinear quantum amplifiers is in its infancy.
We now briefly summarize the early work in the field.
\citet{Bond93} proposed a nonlinear phase-insensitive amplifier with no additive noise. This allows construction of a receiver that saturates the Helstrom limit for binary discrimination of coherent states, using a conceptually simpler approach than the earlier proposal by~\citet{Dolinar73}.
Kouznetsov and coworkers \cite{KouzCot94,*KouzOrte95,*KouzOrteRohr95,*KouzRohr96,*KouzRohr97}, derived general bounds on the noise added by nonlinear amplifiers and found that the added noise could be less than that of a linear amplifier.
They also derived specific approximate input-output relations for nonlinear amplifiers. In a more concrete approach, a number of authors have proposed input-output relations for nonlinear amplifiers specific to photon counting \cite{Yuen86,*Yuen86b,*DAriano92a,*DAriano92b,*HoYuen94,*DAriSaccYuen99,EnkProp19}. So far this early work on nonlinear amplification has received little attention.

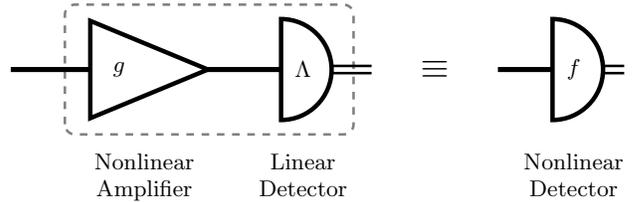
\begin{figure}[t!]
\resizebox{0.99\columnwidth}{!}{
\begin{tikzpicture}
\draw[line width=1pt,draw=gray,dashed, rounded corners] (-.75,-0.9) rectangle (3.25,0.9);
\draw[line width=2pt] (-1.5,0) -- (-0.4,0);
\draw[line width=2pt] (1.2,0) -- (2.25,0);
\node [isosceles triangle, draw=black,line width=1.75pt, minimum width=1.35cm] (0,0) {$g$}; 
\draw (0.35,-1.5)  node[font=\small] { \begin{tabular}{c} Nonlinear\\ Amplifier \end{tabular} };
 \draw (2.55,0) node[semicircle, draw=black,line width=1.75pt, minimum width=1.4cm, shape border rotate=270] {$\Lambda$};
\draw (2.55,-1.5)  node[font=\small] { \begin{tabular}{c} Linear\\ Detector \end{tabular} };
\draw[line width=1.0pt] (2.95,0.05) -- (3.5,0.05);
\draw[line width=1.0pt] (2.95,-0.05) -- (3.5,-0.05);
\draw (4.375,0)  node[font=\Large] {$\equiv$};
\draw[line width=2pt] (5.25,0) -- (6,0);
\draw[line width=1.0pt] (6.7,0.05) -- (7,0.05);
\draw[line width=1.0pt] (6.7,-0.05) -- (7,-0.05);
 \draw (6.3,0) node[semicircle, draw=black,line width=1.75pt, minimum width=1.4cm, shape border rotate=270] {$f$};
 \draw (6.3,-1.5)  node[font=\small] { \begin{tabular}{c} Nonlinear\\ Detector \end{tabular} };
\end{tikzpicture}
}\vspace{-9pt}
\caption{The triangle denotes a nonlinear amplifier with input-output relation $\aout = g \fin +\bin$, where $\fin$ is a normal operator that is nonlinear in $\ain$ and/or $\ain^\dag$, and $g$ is the gain.  We show that in the high-gain limit, a measurement of $\fin$ may be implemented via application of this amplifier followed by an inefficient linear measurement $\Lambda$. 
The simplest nontrivial case is $f=a^\dag a$, which implements a photon number detector.
The internal mode $\bin$ of the amplifier is not depicted. 
}\label{fig1}
\end{figure}

The purpose of this article is to systematically develop the basic notions of quantum nonlinear amplifiers. Our presentation is guided by the realization that nonlinear amplification may be viewed as a kind of transduction between different observables in the same system or between observables in different systems. For example, nonlinear quantum amplifiers may transduce some observable that is nonlinear in the field modes into a linear quadrature of the field modes, as illustrated schematically in \cref{fig1}.

In \Cref{sec:linamp}, we briefly review the established theory of phase-preserving quantum linear amplifiers.
In \Cref{sec:gen_non_lin_amp}, we develop a class of two-mode amplifiers that amplify any normal operator (operators that commute with their adjoint).
For these amplifiers the added noise at the output can be less than {\em a half-quantum of vacuum fluctuations, independent of gain}.
Moreover these amplifiers can facilitate measurement of the normal operator with a noisy linear detector, as illustrated in \cref{fig1}.
Since the class of normal operators includes both Hermitian and unitary operators, this constitutes a practical route to ideal measurements of a broad range of operators. We present explicit Hamiltonians generating these amplifiers; such interactions generalize the von Neumann model of measurement.  
We illustrate the general formalism with examples, including a photon number resolving detector built from a nonlinear amplifier, which show an improvement of measurement efficiency over linear amplification techniques.

In \Cref{sec:three_mode_non_lin_amp} we introduce a class of three-mode nonlinear amplifiers that also amplify normal operators. In these amplifiers the amplified output signal is encoded into a system of two meters, which is similar to the famous model of  \citet{ArthursKelly1965}. This realization of the normal operator nonlinear amplifier provides a further improvement in the noise properties over the two mode realization. 

In \Cref{sec:singlemode} we discuss single mode nonlinear amplifiers.
These amplifiers enable measurement of any function of a field quadrature of that mode. Such amplifiers also require some single mode squeezing. The added noise in these amplifiers is $O(e^{-r})$, where $r$ is the squeezing factor, so the noise may be exponentially decreased by increasing the squeezing.
We conclude with a summary and discussion of some open questions in \Cref{sec:conclusion}.

\section{Phase-preserving linear amplification}\label{sec:linamp}
Real measurement devices are inherently noisy. 
It is thus common to amplify signals prior to measurement. 
A quantum amplifier works by transforming some combination of the ``input'' mode operators unitarily to ``output" mode operators in a way that facilitates more precise measurements by subsequent noisy detection.
In this section we briefly review the theory of quantum linear amplifiers, established in the work of \citet{HausMull62}, generalized and refined by \citet{Cave82}. Readers seeking a detailed general introduction to this subject are encouraged to consult references~\cite{HausMull62,Cave82,ClerDevoGirv10,CaveCombJian12, Aumentado2020}.

Consider a signal carried in a single bosonic mode $\ain$, obeying $[\ain,\ain\dg]=1$. 
In the Heisenberg picture, the amplifier acts to transform $\ain$ into $\aout= U^\dag \ain U$.\footnote{We work in a frame in which fast oscillations at the carrier frequency $e^{\pm i\omega_c t}$ are removed for all modes.}
The simplest amplifiers are the linear amplifiers with input-output relations 
\begin{equation}
\aout = \tau_0 \ain + \tau_1 \ain\dg + \tau_2 \bin + \tau_3 \bin\dg \label{eqn:twomode},
\end{equation}
where $\bin$ is an internal mode of the device.
Unitarity of this transformation is equivalent to the condition that $\aout$ satisfies the bosonic commutation relation, which constrains the coefficients $\tau_i$~\cite{Cave82}. Moreover since $\aout$ is linearly related to the input operators, this is a linear transformation on the input operators $\ain$ and $\bin$.

The most widely-studied linear amplifiers are the phase-preserving linear amplifiers, for which the added noise is independent of the phase of the input signal \cite{HausMull62,Cave82}. These are a special case of \cref{eqn:twomode} with 
\begin{equation}\label{eq:phase_pres_lin_amp}
\aout = g \ain +  \sqrt{g^2 -1} \bin^\dag,
\end{equation}
where $g>1$ is the amplitude gain,  $L^\dag=\sqrt{g^2 -1} \bin^\dag$ is the added noise operator arising from the coupling to the internal mode $\bin$. \cref{eq:phase_pres_lin_amp} can be realized via a parametric interaction generating the two-mode squeeze operator $S(r) = \exp[r(a b - a\dg b\dg)]$, where the squeezing parameter $r$ is related to the gain by $g = \cosh r$~\cite{CaveCombJian12}. Typically one assumes the initial states in modes $a$ and $b$ are uncorrelated. If $\expt{\bin}=0$, as is the case when the $b$ mode starts in vacuum, the output contains an amplified version of the input: $\expt{\aout}= g \expt{\ain}$~\cite{Cave82}.

The noise of an amplifier is defined to be the symmetrically ordered second moment of the operator carrying the signal 
\begin{equation}\label{eq:linearnoise}
\expt{ |\Delta \aout|^2 } =g^2 \expt{ |\Delta \ain |^2 } + (g^2 -1)\expt{ |\Delta \bin|^2 }, 
\end{equation}
where for some operator $O$, $\Delta O\defeq O -\expt{O}$ and $|O|^2\defeq \half ( O O\dg + O\dg O)$.  The first term of \cref{eq:linearnoise} represents the amplified noise intrinsic to the signal and the second term represents the noise added by the amplifier. Applying the uncertainty principle to the added noise operator gives~\cite{Cave82}
\begin{equation}
\expt{|\Delta L|^2}\ge \half |\expt{[L,L^\dag]}| = \half (g^2-1)\,,
\label{eqn:addednoise}
\end{equation}
which implies that for non-trivial gain, i.e., for $g > 1$, the added noise on the bosonic mode is bounded below by $1/2$ in units of number quanta. 
Note also that the added noise is independent of the input signal, but does depend on the gain.

The symmetric variance of $a$ written in terms of quadratures, is $\langle \modulus{\Delta a}^2 \rangle=\frac{1}{2}[ \langle(\Delta x)^2\rangle+ \langle(\Delta p)^2\rangle]$. Thus we may write \cref{eq:phase_pres_lin_amp} in terms of Hermitian quadrature components  $\xout$ and $\pout$ that are related to the modal annihilation operator by $\aout=(\xout +i\pout)/\sqrt{2}$ and obey $[\xout,\pout]=i$. The added noise with respect to these quadrature components is 
\begin{subequations}\label{eq:la_quad}
\begin{align}
\expt{ (\Delta \xout) ^2 }  &= g^2\expt{(\Delta \xin)^2}+  (g^2 -1)\expt{(\Delta \xbin) ^2}\\
\expt{ (\Delta \pout )^2 }  &= g^2\expt{(\Delta \pin)^2}+  (g^2 -1)\expt{(\Delta \pbin)^2},
\end{align}
\end{subequations}
and the added noise is termed phase-insensitive if $ \expt{(\Delta \xbin)^2}=\expt{(\Delta \pbin)^2}$. 

One of the main uses of quantum amplification is in measurement. A general quantum measurement is represented by a positive operator valued measure (POVM) whose elements correspond to measurement outcomes~\cite{nielsen2002quantum}. In practice, there is often a distinction between the desired ideal POVM and the one that represents the noisy measurement performed in the lab. This noisy measurement may be represented by a ``smeared" version of the ideal POVM elements. It is possible, however, to approach the ideal measurement by applying an appropriate amplifier to the system before use of the noisy detector \cite{DallDAriSacc10,Caves20_su11}. 
In particular, for the case of a phase-preserving linear amplifier, a noisy heterodyne measurement can be purified to an ideal measurement in the large gain limit~\cite{HausMull62,Cave82} as shown in detail by ~\cite{DallDAriSacc10}. 
In the following section we shall establish a similar result for nonlinear amplifiers.

\section{Two-mode Nonlinear amplifiers}\label{sec:gen_non_lin_amp}
In this section we show how to transduce a signal carried by an operator which may be nonlinear in the input creation and/or annihilation operators into a linear observable, at the cost of adding exactly a half quantum of vacuum noise. 

We consider the class of amplifiers with input-output relation
\begin{equation}\label{nonlinamp}
\aout = g \fin + \bin,
\end{equation}
where $\fin$ is a normal operator on the Hilbert space of mode $a$, $g$ is the gain, and $b$ is an internal mode. 
Such an amplifier is reasonable to study when mode $a$ is used to carry a signal encoded in the statistics of the operator $f$, which we refer to as the signal operator. The normality of the signal operator ensures that $\aout$ inherits the canonical commutation relation directly from $\bin$.

\Cref{nonlinamp} is deceptively simple.
Recall that a normal operator is one that commutes with its adjoint, $\comm{f}{f^\dag}=0$, and to which the spectral theorem applies \cite{BernauNormal,*BernauUnbounded66}.  For example, when $f$ has a discrete spectrum we have
\begin{align}\label{eq:itsnormal}
    f = \sum_i \mathfrak{f}_i \ket{e_i}\!\bra{e_i}\quad {\rm where} \quad \mathfrak{f}_i \in \mathbb{C},
\end{align}
and $\ip{e_i}{e_j}= \delta _{i,j}$. Thus a measurement of a normal operator can be constructed from projective measurements of $E_i = \ket{e_i}\!\bra{e_i}$~\cite{DunJan13,Roberts2018}. Although \cref{eq:itsnormal} has a discrete spectrum, the results in this manuscript also hold when $f$ has a continuous spectrum. Whenever $\fin$ cannot be written as linear combination of $\ain$ and $\ain^\dag$, we say that the amplifier is nonlinear.
Thus the nonlinear amplifier in \cref{nonlinamp} includes amplification of normal operators such as $a^4 - a^{\dag 4}$, unitary operators such as $e^{-i \chi a\dg a\dg a a }$, and Hermitian operators such as the parity operator $(-1)^{a\dg a}$.

Just as for the linear amplifier, when $\expt{\bin}=0$ the amplified signal is contained in the first moment of the output operator,
\begin{equation}\label{eqn:main_result}
\expt{\aout } = g \expt{\fin}.
\end{equation}
Therefore, when $\expt{\bin}=0$ we can transduce the expectation value of a nonlinear operator into that of a linear operator.

The noise on the output signal may be characterized by the symmetrically ordered second moment of $\aout$,
\begin{equation}\label{eqn:main_result_noise}
\expt{ |\Delta \aout |^2 } 
 =g^2 \expt{|\Delta \fin |^2 }+ \expt{|\Delta \bin |^2}.
\end{equation}
Notice that the added noise is {\em gain independent}, in contrast to the gain dependence of the noise evident in \cref{eq:linearnoise}. Moreover, if we assume that the internal mode $b$ is prepared in the vacuum state, then the added noise is equal to exactly one half quantum of vacuum noise: $\expt{|\Delta b|^2} = \half$\footnote{The noise on the mode is conventionally specified in units of number quanta.}.
Now by decomposing $\fin$ into real and imaginary parts, i.e. $\fin= ( f_R + i f_I)/\sqrt{2} $ with $[f_R, f_I]=0$, we can then determine the noise in the $x$ and $p$ quadratures of $\aout$:
\begin{subequations}\label{eq:nla_quad}
\begin{align}
\expt{(\Delta \xout) ^2} & = g^2 \expt{(\Delta f_R)^2}  + \expt{(\Delta \xbin)^2}\\
\expt{(\Delta \pout)^2} & = g^2 \expt{(\Delta f_I)^2} + \expt{(\Delta \pbin)^2}.
\end{align}
\end{subequations}
We see that the real and imaginary parts of $\fin$ are encoded in orthogonal quadratures of $\aout$ and are amplified equally. This is directly analogous to \cref{eq:la_quad} for the linear amplifier. If the added noise on the two quadratures is symmetric at the input, it will also be symmetric at the output. However, a crucial difference from the linear amplifier is that now the added noise is gain-independent, in both quadratures. 
Finally, \cref{eq:nla_quad} highlights the fact that in order to measure the signal encoded in a normal operator $\fin$, since both $f_R$ and $f_I$ are required, it is necessary to measure both output quadratures. This can be accomplished via heterodyne measurement.

An important special case of  \cref{eqn:main_result,eqn:main_result_noise} is when the signal operator $f$ is Hermitian. In this case no signal is carried in the $\pout$ quadrature of $\aout$ and the signal and noise of the output quadrature $x_{\rm out}$ are given by
\begin{subequations}\label{eqn:main_result_quad}
\begin{align}
\expt{x_{\rm out} } &= \sqrt{2} g \expt{\fin},\\
\expt{ (\Delta x_{\rm out})^2 } 
& = 2 g^2 \expt{(\Delta \fin )^2 }+ \expt{(\Delta \xbin)^2},
\end{align}
\end{subequations}
respectively.
As before, the added noise is independent of gain. When the $\bin$ mode is prepared in vacuum the added noise is equal to $\expt{\Delta \xbin^2} = \half$.\footnote{The added noise on the quadrature is given here in dimensionless units, which translate to $\hbar/\omega$ for electromagnetic modes and to $\hbar/m\omega$ for an oscillator with mass.}  \Cref{eqn:main_result_quad} shows that in this case one can measure the expectation value of the signal operator $\langle \fin \rangle$ by performing a homodyne measurement of the output $x$ quadrature. Note that squeezing the internal mode $b_{in}$ at the input will result in reduced noise $\expt{\Delta \xbin^2} = \half e^{-2r}$ for $r>0$, where $r$ is the squeezing parameter from the single-mode squeezing operator $S(r,\phi) = \exp[r(b^2e^{-2i\phi}-b^{\dag 2}e^{2i\phi})] $~\cite{CaveSchu85a}.

In order to demonstrate the utility of this kind of amplifier, we consider the task of estimating the expectation of a Hermitian operator $\fin$ in a given input state. We apply the amplifier described by \cref{nonlinamp} with mode $b$ in vacuum and then perform a homodyne measurement whose outcome enables estimation of the expectation value. We then define the estimator
\begin{equation}
\widehat{\expt{\fin}} = {\mathbb{E} }[{x_{\rm out}] }/\sqrt{2} g   , 
\end{equation}
where the hat\, $\widehat{}$\, denotes an estimator and $\mathbb{E}[.]$ denotes an expectation over measurement trials. This estimator is unbiased: $ \mathbb{E}\big [\widehat{\expt{\fin}} \big]=\brak{\fin}$. The variance of the estimator quantifies its precision:
\begin{equation}
\textrm{Var}\big [\widehat{\expt{\fin}} \big ]=\textrm{Var}\left[\fin\right]+\frac{1}{4g^2},\label{eq:nonlinearvar}
\end{equation}
where $\textrm{Var}\left[O\right]=\expt{O^2} -\expt{O}^2$ is the variance of a projective measurement of the operator $O$.
Thus, in the large gain limit $g \rightarrow \infty$, this strategy using nonlinear amplification followed by linear (homodyne) detection will yield an estimator with the same variance as a perfect projective measurement of $\fin$.  Furthermore, as noted above, if the internal mode $b$ is prepared in a \textit{squeezed} vacuum state, the added noise may be further reduced at finite gain.

\subsection{Hamiltonian and Unitary realization}\label{sec:meas_model}
In order to aid the realization of amplifiers with the properties described above, we now turn our attention to constructing a physical interaction that gives rise to the input-output relation in \cref{nonlinamp}. We introduce an internal mode $b$ and consider the interaction
\begin{equation}\label{eq:Hint}
H_I=-i\kappa (f^\dagger b-f b^\dagger),
\end{equation}
 where $\kappa $ is a coupling constant, and $f$ and $b$ are operators defined at the input.
 Realizing a large gain in such a Hamiltonian could be achieved by introducing a pump mode in a high-amplitude coherent state, which is thus approximately classical. The unitary generated by this Hamiltonian is $U= \exp(-iH_It) $\,(we set $\hbar=1$ throughout), which results in the following input--output relation on the $b$ mode,
\begin{equation}\label{eq:io_from_Hint}
\bout=U\dg\bin U=g \fin+ \bin,
\end{equation}
with gain $g=\kappa t$. For a detailed derivation of \cref{eq:io_from_Hint}, see Appendix \ref{app:meas_mod_normal}. 
To bring \cref{eq:io_from_Hint} into the form of \cref{nonlinamp}, we assume that $b$ is an oscillator mode and perform a continuous-variable (CV) SWAP gate between modes $a$ and $b$ after the action of  $U$. The CV SWAP exchanges the full Hilbert space of the two modes~\cite{CV_SWAP_Wang_2001}. 

We now turn our attention to the specific case in which $f$ is Hermitian. In this situation \cref{eq:Hint} becomes
\begin{equation}\label{eq:vn_mm}
H_I=\sqrt{2}\kappa f  p,
\end{equation}
where $p$ is the momentum operator on the internal mode $b$. In practice, $p$ may be replaced with any quadrature operator. We shall refer to the unitary generated by $H_I$ as $V$. In \cref{app:meas_mod_hermitian} we show that $V$ induces the correct input-output relation for the $b$ mode.  Performing a SWAP operation between $a$ and $b$ modes after implementing $V$, we arrive at \cref{nonlinamp}.

One of the interesting things about the Hamiltonian constructions of \cref{eq:Hint,eq:vn_mm} is that they are more general than the input-output relation given in \cref{nonlinamp}, because here $f$ need not be a modal operator. 
For example, the $f$ appearing in \cref{eq:Hint} could be a qubit operator such as $f= \op{0}{1} -\op{1}{0}$, which is normal but non-Hermitian. In this case, \cref{eqn:main_result_noise} still applies, with $\aout \leftrightarrow \bout$.
For the Hermitian example of \cref{eq:vn_mm} there are a few instances already recognised and experimentally implemented in the literature. Thus the ``longitudinal interaction'' of references~\cite{Nico2015,Touzard2019, Ikonen2019} can be viewed as a nonlinear amplifier for which $f=\sigma_z$, and the $x$ quadrature is used instead of $p$.  The corresponding interaction Hamiltonian is proportional to $\sigma_z (b+b\dg)$, and thus \cref{eqn:main_result_quad} applies. The same is true of the standard optomechanical interaction Hamiltonian $a\dg a (b+b\dg)$. In this case input amplitude fluctuations in mode $a$ get amplified into output quadrature fluctuations \cite{Kurt94,*LaflammeClerk2011}. This amplification action has been experimentally implemented in Ref.~\cite{ClaLecSimAumTeu16}.

\subsection{Use with noisy linear measurements}\label{sec:meas_purify}
We now show that preamplification by a nonlinear amplifier followed by a \emph{noisy} linear measurement, can result in an ideal (i.e., a projective) measurement of the normal operator $\fin$ in the large gain limit, as depicted in~\cref{fig1}.
Linear measurements in the optical and microwave domains are typically destructive. For this reason, we focus on determining the POVM elements\footnote{Moreover, these also lead to the canonical Kraus operators, which can be taken as the square root of the POVM elements.}. 

We noted previously that in the large gain limit, the first and second moments of the amplified  signal for  Hermitian $f$ match the statistics of an ideal projective measurement of $f$. This suggests that these types of amplifiers are intimately related to the von Neumann measurement model. Indeed, \cref{eq:vn_mm} {\em is} the classic von Neumann measurement model (see Chap. VI, Sec. 3 in Ref.~\cite{vonNeu2018}\footnote{Specifically ``For this we choose the particular form $\frac{h}{2\pi i } q \frac{\partial}{\partial r}$'' which is $H_{\rm int } = \hbar xp$ in modernized notation.}). Note that in this section we take the amplifier unitaries from \cref{sec:meas_model}, i.e., without making a CV SWAP between the two modes, which means that the output signal is in the $b$ mode and thus the linear measurement is performed on that mode. 

In the case when $\fin$ is normal, it is transduced with gain into mode $b$. When this amplification is followed by a noisy heterodyne measurement of mode $b$, it results in an effective POVM on mode $a$ with elements
\begin{align}
E_\beta 
= \bra{0_b}  U^{\dg} M_{\beta} U \ket{0_b},
\end{align}
where $\beta$ is the complex  measurement outcome and $U$ is the unitary generated by \cref{eq:Hint}, which depends on the amplifier gain $g=\kappa t$.
In Reference~\cite{DallDAriSacc10} it is shown that inefficient heterodyne detection corresponds to the POVM
\begin{equation}
M_{\beta}=\frac{1}{\pi^2\sigma^2} \int d^2\gamma\,  e^{-|\gamma -\beta|^2/\sigma^2} \op{\gamma}{\gamma},
\end{equation}
where $\ket{\gamma}$ is a coherent state, $\sigma^2 = (1-\eta)/\eta$, and $\eta$ is the detector efficiency. This may be interpreted as a Gaussian smearing of the perfect heterodyne POVM element \cite{Caves20_su11}. Using the spectral representation of $\fin$, see \cref{eq:itsnormal}, detailed calculations that are given in \cref{app:purfied_meas_normal_f} lead to the POVM\footnote{ This result is for the case where $\fin$ has a discrete spectrum. To obtain the continuous spectrum case one replaces the sum over $i$ with an integral. }
\begin{align}
\label{eq:booyah}
E_{\beta}(\sigma,g)
&=\sum_i\frac{g^2}{\pi(\sigma^2+1)}\exp\left[-g^2\frac{\modulus{\beta-\mathfrak{f}_i}^2}{\sigma^2+1}\right]\ket{e_i}\bra{e_i}.
\end{align}
\Cref{eq:booyah} describes complex valued outcomes $\beta$ that are normally distributed about the discrete eigenvalues $\mathfrak{f}_i$ of $\fin$.
The complex plane may be tiled with decision regions $R_i\ni \mathfrak{f}_i$, and the outcome $\beta$ associated to the label $i$ for which $\beta\in R_i$ (see Chapter 2 of Ref.~\cite{vanTreesI2004}). 
In the limit $g\rightarrow\infty$, the function inside the sum in \cref{eq:booyah} limits to $\delta(\beta-\mathfrak{f}_i)$, so in this limit the decision regions turn the POVM into a projective measurement of $f$:
\begin{align}
\lim_{g\rightarrow\infty}\int_{R_i} d^2\beta\,E_\beta(\sigma,g)=\ketbra{e_i}{e_i}.
\end{align}
Furthermore, outside of this limit, the signal-to-noise ratio (${\rm SNR}$) increases with gain. For example if the primary mode is prepared in an eigenstate $\ket{e_i}$, then the signal-to-noise ratio is ${\rm SNR} = |\mathfrak f_i|/ \sqrt{(\sigma^2+1)/g^2}$, and thus for any $g>1$ the SNR will then be correspondingly larger than the $g=1$ value.

Now consider the case when $\fin$ is Hermitian and diagonal in the orthonormal basis $\ket{i}$. We apply the amplifier unitary $V$ that is generated by the Hamiltonian in \cref{eq:vn_mm}, and then perform an inefficient or noisy homodyne measurement on mode $b$ with POVM elements
\begin{align}\label{eq:noisyhomapp}
M_x&=\frac{1}{\sqrt{\pi}\sigma}\int dy\,e^{-(y-x)^2/\sigma^2}\ketbra{y}{y},
\end{align}
where $\ket{y}$ is a quadrature eigenstate and $\sigma^2 = (1-\eta)/4 \eta$
~\cite{DallDAriSacc10,Caves20_su11}. 
Mode $b$ is prepared in a Gaussian position wave function centered at $x=0$ 
\begin{align}\label{eq:gaussian_wf}
\ket{\phi}_b = \int dx\, \phi(x)\ket{x}_b \,\,\, {\rm where}\,\,\, \phi(x) = \frac{e^{-x^2/2\epsilon^2}}{(\pi \epsilon^2)^{1/4}},
\end{align}
here the position variance of this wavefunction is ${\rm Var}[x] = \half \epsilon^2$.
In this case, preamplification followed by a homodyne measurement results in an effective measurement of $f$ with POVM elements
\begin{align}
E_x(\sigma,g,\epsilon) 
&= \bra{\phi}_b V^\dag M_xV\ket{\phi}_b \nonumber\\
&=\sum_i\frac{g}{\sqrt{\pi(\sigma^2+\epsilon^2)}}\exp \left [-g^2\frac{(x-f_i)^2}{\sigma^2+\epsilon^2} \right ]\ketbra{i}{i}.
\end{align}
The case of $\epsilon^2=1$ corresponds to the probe being prepared in vacuum. Then as the probe wave function becomes squeezed in the position quadrature, the contribution from $\epsilon^2=e^{-2r}$ becomes negligible as the squeezing parameter $r$ increases from zero.
Once again, in the large-gain limit, the POVM elements approach an ideal projective measurement of $\fin$, when the outcomes $x$ are properly coarse-grained into decision regions. 

For the case of $\fin=a^\dag a$, the 
implementation of an effective projective measurement of $\fin$ via this approach of combining preamplification with a noisy measurement has been realized in an optomechanical system~\cite{ClaLecSimAumTeu16}. Figure 4 of Ref.~\cite{ClaLecSimAumTeu16} shows that as the gain increases the effective measurement of the photon number, via an inefficient homodyne measurement, approaches the ideal limit. 
For the case of  $\fin=\sigma_z$, the improvement of measurement quality has been observed in two experiments using superconducting qubits coupled to microwave cavities. In these experiments, the longitudinal interaction~\cite{Nico2015} is used to implement an amplifier that allows higher SNR (see Fig. 2 (b) of Ref.~\cite{Touzard2019}) or lower average measurement error (see Fig. 3 of Ref.~\cite{Ikonen2019}) for qubit readout relative to the usual dispersive measurement accomplished via noisy heterodyne detection.

\subsection{Example: Quadratic amplifiers}\label{sec:quad_amp}
Suppose that we are interested in estimating second moments of $a$, i.e. the quantities  $\brak{a^2}$, $\brak{a^\dag a}$, and $\brak{a^{\dag 2}}$. 
One way to obtain an estimate of this signal is to calculate higher order moments of $a$ from the outcomes of linear measurements, and from this data to infer the second moments. Such techniques can be inefficient, since bounding errors on the estimates of second moments requires the computation of fourth moments (see for example  \cite{corr1}).
An alternative is to use a nonlinear amplifier followed by a linear detector. We now consider the family of such amplifiers for which the signal operator $\fin$ is quadratic in $a$ and $a^\dag$, and analyze the effective measurements resulting from following these amplifiers with linear measurements.

Quadratic amplifiers are described by signal operators of the general form\footnote{One might expect the anti-normally ordered term $aa^\dag$ to appear in \cref{eq:quadratic_f_in}, using the commutation relations one can see that $a a^\dag = a^\dag a + \Id$ which re-scales the $\beta$ term and adds the $\delta$ term.}
\begin{align}\label{eq:quadratic_f_in}
\fin = \alpha a^2+\beta a^\dagger a+\gamma a^{\dag 2} +\delta \Id,    
\end{align}
where $\alpha,\beta,\gamma, \delta \in \mathbb{C}$ and normality of $\fin$ is enforced (see \cref{apdx:twomodecom}) by asserting $|\alpha|^2=|\gamma|^2$, $\alpha \beta^* = \beta \gamma^*$, and $\delta$ is unconstrained.
Notice in \cref{eq:quadratic_f_in} we have dropped the `in' designation on the modal operators for notational compactness. Thus, as shown in the previous section, these amplifiers, used in conjunction with linear detectors, can measure operators quadratic in the field modes with half a quantum of added noise.

We now examine two simple cases of the quadratic amplifiers given in \cref{eq:quadratic_f_in}. These are the signal operators
\begin{align} \label{eq:quad_amps}
 f_\pm = \pm \half ( a^2 + a^{\dag 2}) +a^\dag a + \half \Id .
\end{align}
It is easy to verify that the signal operator $f_+$ corresponds to $x^2$. Then \cref{eqn:main_result_quad} tells us that the operator $x^2$ is transduced into the $\xout$ quadrature, which implies that the $n$th moment of $\fin$ is transduced into the $n$th moment $\xout$. 
Because $x^2$ is Hermitian, the added noise may again be suppressed by squeezing the internal mode $b$ of the amplifier as discussed above.  Similarly, the $f_-$ signal operator is seen to correspond to $p^2$, and hence $p^2$ is transduced into $\xout$ by the $f_-$ amplifier.

Photon number amplifiers have applications in microwave and optical photon number detection.
To obtain a photon number amplifier $f_n$ from  \cref{eq:quadratic_f_in}, we choose $f_n \equiv \fin  = a^\dagger a $. 
In the optical domain, significant experimental progress has been made on number resolving detectors and their ultimate limitations~\cite{YounSaroFran2018,YangJacob2020}. Nevertheless, they remain costly and inefficient.
In the microwave domain, much theoretical~\cite{Helmer2009, *GoviaPritchett2014, *Bixuan2014, *Schondorf_2018, *grimsmo_quantum_2020} and experimental~\cite{Kono2018,*Opremcak2018,*Besse2018,*Lescanne2020} effort has gone into developing number resolving detectors for itinerant  microwave  photons. For this reason we investigate this amplifier in more detail below. 

If one assumes a Fock state  $\ket{n}$ in the input mode $\ain$ and vacuum or squeezed vacuum in the $\bin$ mode, then from \cref{eqn:main_result_quad} we have
\begin{align}\label{eq:not_special}
\expt{\xout} = \sqrt{2}g n \quad {\rm and} \quad \expt{|\Delta \xout |^2} = \half \,\, {\rm or} \,\, \half e^{-2r},
\end{align}
which implies a signal-to-noise ratio ${\rm SNR} = \expt{\xout} / \sqrt{\expt{|\Delta \xout |^2}}$ of $2gn$ for the vacuum input and $2e^r gn$  for a squeezed input. Thus with this amplifier and a linear detector, one can measure the photon number in a single shot when $g\gg 1$.\footnote{Making the gain more nonlinear in the photon number could be even more advantageous, e.g., with $f_M = (a\dg a)^M$, the SNR becomes ${\rm SNR} = 2gn^M$ when $\bin$ is in vacuum.} 
This number amplifier is related to but distinct from earlier work \cite{Yuen86,*Yuen86b,*DAriano92a,*DAriano92b,*HoYuen94,*DAriSaccYuen99,EnkProp19} that has proposed photon number amplifiers with input-output relations of the form $\aout\dg\aout \propto g \ain\dg \ain$, i.e., input photon number is transduced into output photon number. 
The primary difference between our nonlinear number amplifier and these other proposals is that our signal is stored in a field quadrature. This permits a number resolving measurement to be implemented using linear detectors, which is easier to implement than with existing number resolving detectors. Recently an input-output relation $\aout = g n \otimes \op{\psi_0}{\psi_0} + \bin$ was proposed in Ref.~\cite{BiswasvanEnk2020}. This differs from our input-output relation \cref{nonlinamp} by the further inclusion of a projector $\op{\psi_0}{\psi_0}$ onto a third system with Hilbert space $\mathcal H_c$.

We now compare the statistical efficiency of estimating the mean photon number of a signal using our nonlinear amplifiers to what can be achieved using linear amplifiers. 
Suppose that instead of a nonlinear amplifier and a linear detector as considered in the previous subsection, we have a phase-preserving linear amplifier, see \cref{eq:phase_pres_lin_amp}, together with a linear detector. We consider the linear amplifier to be applied to the mode $a$ with the internal mode $b$ in vacuum, followed by a heterodyne measurement of mode $a$, resulting in the outcome $\alpha \in \mathbb{C}$. We define the estimator for the mean photon number as
\begin{align}
\hat{n}&=\frac{|\alpha|^2}{g^2}-1.
\end{align}
Computing the mean and variance of $\hat{n}$ yields (see \cref{appendix:estimators} for the derivation)
\begin{subequations}
\begin{align}
\mathbb{E}\left[\hat{n}\right]&=\brak{a^\dagger a}\\
\text{Var}\left[\hat{n}\right]&=\text{Var}\left[a^\dagger a\right]+\brak{a^\dagger a}+1.\label{eq:est_var}
\end{align}
\end{subequations}
Here $\text{Var}\left[a^\dagger a\right]$ is the variance of a projective measurement of the photon number operator. Comparing the variance in \cref{eq:est_var} to \cref{eq:nonlinearvar}, we see that for $g>1/2$ the scheme of nonlinear amplification with $f = a^\dagger a$ followed by homodyne detection provides a clear improvement for number measurement over this entirely linear scheme.

\section{Three-mode Nonlinear amplifiers}\label{sec:three_mode_non_lin_amp}

In this section we consider a special class of three-mode nonlinear amplifiers that amplify normal operators, and that have improved noise properties relative to their two mode counterparts in \Cref{sec:gen_non_lin_amp}. 
Detailed calculations for the results below can be found in \cref{app:AK_like}.

In this setup we wish to measure or to amplify a normal operator $f$ on system $a$. To do so we couple the real and imaginary parts of the normal operator $f$ to two independent probe systems or ``meters'' and then subsequently measure both meter systems to simultaneously reveal information about $f_R$ and $ f_I$. 
Inspired by the model of \citet{ArthursKelly1965}, we consider the Hamiltonian  
\begin{equation}
H_{\rm m} = \kappa (f_R p_b + f_I p_c)\label{eq:AKHam}
\end{equation}
where $p_i$ is the momentum operator for the $i$'th internal mode. The unitary transformation generated by $H_{\rm m}$ is denoted by $W$.
This unitary amplifies (or transduces) the real and imaginary parts of $f$ into the position quadratures of the two meters in modes $b$ and $c$, respectively. This is apparent from the input-output relations for the position operator of both meters
\begin{align}
x_{b}^{\rm out} = g f_R + x_b^{\rm in}, \quad {\rm and}\quad 
x_{c}^{\rm out} = g f_I + x_c^{\rm in}, 
\end{align}  
where $g= \kappa t $ as before.
The corresponding noise in the position quadratures at the output is
\begin{subequations}\label{eq:three_mode_nla_quad}
\begin{align}
\expt{(\Delta x_{b}^{\rm out}) ^2} & = g^2 \expt{(\Delta f_R)^2}  + \expt{(\Delta x_b^{\rm in})^2}\\
\expt{(\Delta x_{c}^{\rm out})^2} & = g^2 \expt{(\Delta f_I)^2} + \expt{(\Delta x_c^{\rm in})^2}.
\end{align}
\end{subequations}
These expressions resemble \cref{eq:nla_quad}, the noise in those expressions is either unequally correlated (perhaps due to squeezing) between the two quadratures or equal to the minimal value of $\half$ in both quadratures. In contrast, here we can start with the two auxiliary modes in uncorrelated squeezed states.  This allows us to generate a reduction of the added noise in both meters, specifically $\expt{(\Delta x_i^{\rm in})^2} = \half e^{-2r}$ for $i\in [b,c]$. Thus the total added noise can approach {\em zero} as $r\rightarrow \infty$.

We now show that when supplemented with linear detectors, this two meter measurement  results in a POVM that measures the normal operator $f$.
Since there are two meters in this setup, the heterodyne detection by a single meter in \cref{sec:gen_non_lin_amp} is now replaced with two meters that output the measurements of two noisy homodyne detectors. We prepare the two meters in uncorrelated Gaussian states, see \cref{eq:gaussian_wf}. To derive the expressions below, we have assumed equally noisy homodyne detectors, as characterized by $\sigma^2$, and the position wave functions of meters $\phi_b$ and $\phi_c$  have the same variance.

In this situation the coupled amplification and homodyne measurements result in a POVM on the Hilbert space $\mathcal H_a$ of mode $a$:
\begin{align}
E_{x_b,x_c} 
&= \bra{\phi_b,\phi_c} W^\dag  M_{x_b} M_{x_c} W \ket{\phi_b,\phi_c} \\
&= \sum_{k} \frac{g^2 }{\pi( \sigma^2 + \epsilon^2)} \times \nonumber\\
& \phantom{=}\exp \left [ -g^2\frac{(x_b-\mathfrak{f}_k^R)^2 + (x_c-\mathfrak{f}_k^I)^2}{\sigma^2 + \epsilon^2} \right ]
\op{e_k}{e_k}. \label{eq:intermeidatepovm} 
\end{align}
Here $x_b$ and $x_c$ are the outcomes of the two independent noisy homodyne measurements and $\epsilon^2/2$ is the variance of the position wave function of the meters. Clearly the two measurement outcomes of this POVM are peaked around the real and imaginary eigenvalues values of the normal operator $f$. 

To facilitate a direct comparison of this POVM to the POVM in \cref{eq:booyah}, we change variables and define a single complex outcome from the two real outcomes, $\beta=x_b +ix_c $. Recognizing $\mathfrak{f}_k = \mathfrak{f}_k^R + i \mathfrak{f}_k^I $, we then see that \cref{eq:intermeidatepovm} can be written as
\begin{align}\label{eq:AKmodel}
E_{\beta}(\sigma,g,\epsilon) 
&= \sum_{k} \frac{g^2 }{\pi( \sigma^2+ \epsilon^2)} \exp \left [ -g^2\frac{|\beta -\mathfrak{f}_k|^2 }{\sigma^2 + \epsilon^2} \right ]
\op{e_k}{e_k}.
\end{align}
Now we relate the meter variances to the initial squeezing,  $\epsilon^2=e^{-2r}$. This implies that for $r>0$, the POVM in \cref{eq:AKmodel} has less noise than the POVM in \cref{eq:booyah}.  Nevertheless, in the high-gain limit, both of these POVMs limit to the ideal measurement of the normal operator $f$.

\section{Single mode nonlinear amplifier}\label{sec:singlemode}
In a single mode, a linear transformation from the input annihilation operator to the output operator has the form
\begin{equation}\label{eqn:onemode}
\aout = \mu \ain + \nu \ain\dg,
\end{equation}
where $\mu = \cosh r$ and $\nu = -e^{2i \phi}\sinh r$; c.f. \cref{eqn:twomode}.
A device that implements this transformation is known as a single-mode linear amplifier. The transformation in \cref{eqn:onemode} is enacted by a single-mode squeezing operator \cite{CaveSchu85a}. In some situations this is known as phase sensitive amplification because the gain and noise both depend on the squeezing angle $\phi$, i.e. on which quadratures are squeezed and antisqueezed~\cite{Cave82}. 
In microwave quantum optics, phase sensitive amplifiers are used to turn heterodyne measurements into homodyne measurements by selectively amplifying a desired quadrature. It is well known that phase-sensitive amplifiers can also be used to improve the quality of inefficient homodyne measurement when used as a preamplifier, see e.g. Refs~ \cite{Cave82, OckeloenKorppi2017,KnySpaCheKha2018, LauClerk2019, Caves20_su11, BaiVenuKuns20} and \cite{Jacobsen18,Derkach20, NohGirLia2020}\footnote{In these references the noise channel preceding measurement can be composed with the measurement operators to give an effective noisy POVM that is purified by pre-amplification.}. This is also explained in detail by \citet{DallDAriSacc10}.  

Our goal in this section is to develop single-mode non-linear amplifiers. Such amplifiers could be used to, e.g., coherently transduce $f$ that is non-linear in $\ain$ into a linear quadrature such as $\pout$ for quantum control purposes, or to allow perfect measurements of $f$ with a linear detector.
Because the analysis is necessarily phase sensitive, we define a rotated mode $a_\theta = a e^{i\theta}$ and a rotated quadrature $x_\theta = (a_\theta+a_\theta\dg)/\sqrt{2}$.
For a single mode, we propose nonlinear amplifier input-output relations of the form
\begin{align}
\aout &= \gamma \fin(x_\theta) + \tilde \mu a_\theta + \tilde\nu a_\theta\dg, 
\end{align}
where $f$ is any well behaved function of the rotated quadrature $x_\theta$, and $\gamma =g e^{i\varphi}$, $\tilde \mu = \mu e^{i\theta}$, $\tilde \nu = \nu e^{i\theta}$. When the signal operator $f$ is nonlinear in $x_\theta$, a device that implements this transformation is called a nonlinear amplifier. It has been shown that the unitarity constraint is enforced by ensuring that
$|\tilde \mu|^2-|\tilde \nu|^2=1$ and ${\rm Re}[\tilde \mu \gamma^* - \tilde \nu^* \gamma]=0$~\cite{SienLisi01,*WuCote02,*DellSiena04,*BrunHouh19}. These input-output relations may be realized via Hamiltonian evolution~\cite{SienLisi01,*WuCote02,*DellSiena04,*BrunHouh19} or via measurement-based approaches~ \cite{Konno2020}. 

The amplifier described above contains a phase redundancy. To simplify the analysis, we specialize to $\theta=0$, $\varphi = \pi/2$, $\phi =-\pi/2$, which gives the input-output relation
\begin{align}\label{eq:specialcase_single_mode}
\aout &=  i g \fin ( \xin ) + \cosh r \ain + \sinh r \ain\dg.
\end{align}
If we measure the quadrature $x_{{\rm out}}$, we obtain no information about the signal operator: $x_{\rm out} = e^r x_{\rm in}$. If, however, we look at the $\pout $ quadrature, then we find
\begin{align}
\pout &= \sqrt{2}g \fin(\xin) +  e^{-r} \pin \,.
\end{align}
Examining the second moment of $\pout$ in the large gain limit yields
\begin{align}
\expt{(\Delta \pout)^2} & \approx  2g^2 \big \langle\big(\Delta \fin(\xin)\big)^2 \big \rangle + O( g e^{-r} ),
\end{align}
since $e^{-r}$ and $e^{-2r}$ both tend to zero as $r$ increases (see \cref{app:singlemodeexample} for more details). Thus in the limit that $r\rightarrow \infty$, these amplifiers can i) transduce $\fin(\xin)$ into the linear quadrature $p$, and ii) allow perfect measurements of $\fin(\xin)$ with a linear detector,  without adding noise. The simplest nontrivial case is $\fin(x)=x^2$, which would allow measurement of $x^2$ without obtaining any information about $\textrm{sgn}(x)$.

\section{Conclusions}\label{sec:conclusion}
Dramatic improvement in linear amplifier performance has significantly advanced quantum technology by facilitating precise measurements of microwave circuits and quantum mechanical oscillators. In this work, we have introduced classes of non-linear amplifiers that have better noise properties than linear amplifiers.
A critical aspect of the analysis and results obtained here is their illumination of the beneficial role of nonlinear amplification in the quantum theory of measurement. We hope that implementation of detectors based on these principles will enable our ``grubby, classical hands''~\cite{Cave82} to access more of the pristine quantum realm.

There are four main achievements of this work. First, we have introduced  a large class of nonlinear amplifiers that have negligible added noise. Second, we have provided a unitary interaction model for normal operators, extending von Neumann's measurement model for Hermitian operators. Third, we have demonstrated that these nonlinear amplifiers, when coupled with linear measurement devices, can achieve better statistical efficiency at estimating signals stored in nonlinear mode observables than is achievable with linear amplification and measurement. Finally, we have shown that these nonlinear amplifiers, when coupled with noisy linear measurements, can nevertheless achieve perfect measurement of normal operators in the large-gain limit. This may find application in fault tolerant measurements of stabilizers and logical operators in bosonic codes with translation~\cite{GKP2001} and rotation~\cite{GCB2020} symmetries.

There are several key insights gained from these results. We take the point of view that amplification or transduction is a perspective through which one can analyze any quantum channel, rather than a property of a particular class of channels.
Moreover is fruitful to think of amplification in terms of transduction, i.e., translating a signal stored in one physical quantity to a signal stored in another. An example of this is the transduction of photon number into a quadrature. 
Another observation is that  preservation of the commutation relations from input to output is equivalent to the existence of a unitary operation that implements our nonlinear amplifiers. If the signal operator is normal, then the commutation relations are enforced entirely by the internal mode. As noted in \cref{sec:meas_model}, the unitary realization given in this work does not require the signal operator $\fin$ to be bosonic. It could alternatively be a spin operator such as $\sigma_z$ or $J_z^2$.  
Lastly, our work elevates normal operators to the status of observables by providing an explicit measurement models for them. The measurement models that we have introduced generalize the von Neumann~\cite{vonNeu2018} model for Hermitian operators, and the Arthurs-Kelly model \cite{ArthursKelly1965} for simultaneous position and momentum measurements.
Thus the class of operators that defines projection-valued measurements is now seen to coincide with the class that can be measured using the coupling to a single (or two) oscillator modes equipped with linear measurement devices.

There are many open questions remaining.  
In the current work we have only explored two specific kinds of nonlinear input--output relation. However, there are many other classes of nonlinear amplifiers that are also conceptually simple. For example, in a single mode one could sensibly consider a Kerr nonlinearity to be a nonlinear amplifier~\cite{KitaYama86,Sasaki1995}. For multiple modes the possibilities are endless. For example, \citet{Bond93} constructs the input-output relation $a_{\rm out} = [a e^{i g c\dg c} + b e^{i\phi}]/\sqrt{2} $, where $a,b$, and $c$ are bosonic operators. This can be easily generalized to $\aout = (a \fin(c,c\dg) + b)/\sqrt{2}$, where $\fin$ is a unitary operator.

Perhaps the most important topic for future work on nonlinear amplifiers is to analyze common experimental nonlinearities and determine what measurements they can enable. We note that many amplifiers in the microwave domain are operated in a linearized regime~\cite{Boutin17, Sivak2019}. It is interesting to analyze such devices in the nonlinear regime~\cite{LiuChien2020}, as well as other nonlinear Hamiltonians, to see what signal operators the devices are capable of measuring outside their linear mode of operation.

{\em Acknowledgments:} The authors acknowledge helpful discussions with Carl Caves, Leigh Martin, C. Jess Riedel, and John Teufel. JE was supported by the National Defense Science and Engineering Graduate (NDSEG) fellowship and by UC Berkeley via startup funding for Prof. Kranthi Mandadapu.  JC was supported by the Australian Research Council through a Discovery Early Career Researcher Award (DE160100356) and by the University of Colorado Boulder.

%

\vspace{2em}

\newpage\newpage
\onecolumngrid
\appendix

\section{Unitary realization}\label{app:meas_mod}
To realize our nonlinear amplifiers we consider the interaction Hamiltonian between the system of interest, with  Hilbert space $\mathcal{H}_a$, and a one-dimensional pointer system described by the Hilbert space $\mathcal{H}_b=L^2(\mathbb{R})$.

\subsection{\texorpdfstring{$f$}{TEXT} is normal}\label{app:meas_mod_normal}
For some normal operator $f$ on mode $a$ we take the interaction Hamiltonian $H_I=-i\kappa (f^\dagger b-f b^\dagger)$,
where $[b,b\dg]=1$. 
This generates a unitary $U(t)=\exp(-iH_It)$, that we order via the Zassenhaus formula
\begin{align}
U(t)
= e^{-iH_It}=e^{-(f^\dagger b-f b^\dagger)\kappa t}
=e^{g f b^\dagger }e^{-g f^\dagger b }e^{-\frac{g^2}{2}\comm{-fb^\dagger}{f^\dagger b}}
=e^{g f b^\dagger}e^{- g f^\dagger b }e^{-\frac{g^2}{2}f^\dagger f},\label{eq:normalU}
\end{align}
we have also defined $g = \kappa t$. The series terminates because $\comm{f^\dag b}{-fb^\dag}=-f^\dag f$ and both $f$ and $f^\dag$ commute with $f^\dag f$.
The auxiliary mode operator $b$ evolves as follows:
\begin{align}
\bout=U^\dag(t)b U(t)
&=e^{-\frac{g^2}{2}f^\dag f}e^{-gfb^\dag}e^{gf^\dag b}\left(\comm{b}{ e^{gf b^\dag}}+e^{gf b^\dag}b\right)e^{-gf^\dag b}e^{-\frac{g^2}{2}f^\dag f}\\
&=e^{-\frac{g^2}{2}f^\dag f}e^{-gfb^\dag}e^{gf^\dag b}\comm{b}{ e^{gf b^\dag}}e^{-gf^\dag b}e^{-\frac{g^2}{2}f^\dag f}+b\\
&=e^{-\frac{g^2}{2}f^\dag f}e^{-gfb^\dag}e^{gf^\dag b}gf e^{gf b^\dag}e^{-gf^\dag b}e^{-\frac{g^2}{2}f^\dag f}+b\\
&=gf+b
\end{align}
Then if a SWAP operation is performed between the $a$ and $b$ modes the input-output relation is exactly \cref{nonlinamp}.

\subsection{\texorpdfstring{$f$}{TEXT} is Hermitian}\label{app:meas_mod_hermitian}
Taking the coupling Hamiltonian ${H}=\sqrt{2}\kappa f_a\p_b$, with $\p_b$ the momentum operator $-i\hbar\partial_x$ on $\mathcal{H}_b$. This is interesting because it is the von Neumann model of measurement, which realizes measurement of a self-adjoint operator $f_a$ via an interaction between the system and a 1D pointer system. We take $\x_b$ to be the position operator on the pointer Hilbert space $\mathcal{H}_b$ and define $V = e^{-i{H}t/\hbar}$.

We will make use of the following identity for two operators $X$ and $Y$ that commute with the commutator $C: = [X,Y]$, then
\begin{equation}
    e^{sX} Y e^{-sX} = Y + s [ X, Y ] ~.
\end{equation}
To compute the input-output relations in our application we identify $X =  i f_a\otimes p_b$, $Y = \id \otimes x_b$ and $s= gt = \sqrt{2}\kappa t $. Thus
\begin{subequations}
\begin{align}
   [X,Y] &= i [ f_a\otimes p_b, \id \otimes x_b] = f_a \otimes I = C\\
   [C, X] &= i[f_a \otimes I, f_a\otimes p_b] = 0\\
   [C, Y] &= [f_a \otimes I,  \id \otimes x_b] = 0.
\end{align}
\end{subequations}
Then we find
\begin{align}
x_b^{\rm out}=V^\dag\x_b V = \x_b+\sqrt{2}g f_a.
\end{align}
Since $H$ commutes with $\p_b$, so we have $V^\dag\p_bV=\p_b$. It is easy to find input-output relation for the annihilation operator:
\begin{align}
a_{b}^{\rm out}=V^\dag {a}_b V&=\frac{1}{\sqrt{2}}V^\dag(\x_b+i\p_b)V=\frac{1}{\sqrt{2}}(\x_b+\sqrt{2} g f_a+i\p_b))= {a}_b+ g f_a.
\end{align}

\vspace{-2pt}

\section{Purified POVM}\label{app:purified_povm}
The purpose of this section is to show that the combination of a nonlinear amplifier and linear measurement results in an ideal measurement of the operator $\fin$ in the large gain limit.

\subsection{\texorpdfstring{$\fin$}{TEXT} is normal and the linear measurement is heterodyne}\label{app:purfied_meas_normal_f}
An ideal heterodyne measurement with complex outcomes $\gamma$ is described by the POVM elements
\begin{equation}
  E'_\beta = \frac{1}{\pi} \op{\gamma}{\gamma},
\end{equation}
where $\gamma$ labels the complex outcome and $\ket{\gamma}$ is a coherent state. Consider preparing a mode $b$ in the vacuum state $\ket{0_b}$, performing the unitary transformation $U$ generated by the Hamiltonian Eq. \eqref{eq:Hint} to the joint system, and performing ideal heterodyne measurement on the $b$ mode. The POVM element associated with the outcome $\gamma$ is
\begin{align}
\Upsilon_\gamma=\bra{0_b}  U^{\dg}E'_\gamma U \ket{0_b}  
&=\frac{1}{\pi}e^{gf \gamma^*+gf^\dagger\gamma }e^{-g^2f^\dagger f}e^{-\gamma^*\gamma}\\
&=\frac{1}{\pi}\sum_i\exp\left[-\big(\gamma^*\gamma-g(\mathfrak{f}_i\gamma^*+\mathfrak{f}_i^*\gamma)+g^2\mathfrak{f}_i \mathfrak{f}_i^*\big)\right]\ket{e_i}\bra{e_i}\\
&=\frac{1}{\pi}\sum_i e^{-\modulus{\gamma-g\mathfrak{f}_i}^2}\ket{e_i}\bra{e_i}.
\end{align}
Note that this is an operator on the $a$ mode. Above we used Eq. \eqref{eq:normalU} to find $U\ket{0_b} =e^{g f b^\dagger} \ket{0_b}e^{-\frac{g^2}{2}f^\dagger f}$ and the spectral theorem  representation of $f$ see \cref{eq:itsnormal}.

Consider a noisy heterodyne measurement, see \citet{DallDAriSacc10}, given by the POVM elements
\begin{equation}
M_{\beta}=\frac{1}{\pi\sigma^2} \int d^2\gamma\,  e^{-|\gamma -\beta|^2/\sigma^2} E'_\gamma =\frac{1}{\pi^2\sigma^2 } \int d^2\gamma  e^{-|\gamma -\beta|^2/\sigma^2} \op{\gamma}{\gamma} , \quad {\rm where} \quad  \sigma^2 :=\sigma^2(\eta) = \frac{1-\eta}{\eta},
\end{equation}
where $\eta$ is the efficiency of the heterodyne detection. Exploiting linearity, we have the effective POVM elements

\begin{align}
E_\beta
=\bra{0}_b U^\dag M_\beta U\ket{0}_b
&=\frac{1}{\pi\sigma^2} \int d^2\gamma\,  e^{-|\gamma -\beta|^2/\sigma^2} \Upsilon_\gamma\\
&=\frac{1}{\pi^2\sigma^2}\sum_i\int d^2\gamma\, e^{-\modulus{\gamma-\beta}^2/\sigma^2}e^{-\modulus{\gamma-g \mathfrak{f}_i}^2}\ket{e_i}\bra{e_i}\\
&=\frac{1}{\pi^2\sigma^2}\sum_i\int d^2\alpha\, e^{-\modulus{\alpha}^2/\sigma^2}e^{-\modulus{\alpha +\beta-g \mathfrak{f}_i}^2}\ket{e_i}\bra{e_i}\\
&=\frac{1}{\pi^2\sigma^2}\sum_i\frac{\pi}{1+1/\sigma^2}\exp\left[-\frac{\modulus{\beta-g \mathfrak{f}_i}^2}{1+\sigma^2}\right]\ket{e_i}\bra{e_i}\\
&=\sum_i\frac{1}{\pi(1+\sigma^2)}\exp\left[-\frac{\modulus{\beta-g \mathfrak{f}_i}^2}{1+\sigma^2}\right]\ket{e_i}\bra{e_i} 
\end{align}
using the convolution of two complex Gaussians.

If we now change variables by dividing the outcome $\beta$ by the gain $g$ to obtain an outcome $\phi=\beta/g$ (the associated change of measure is $  d^2\phi =g^2 d^2\beta$). With respect to this new variable the corresponding POVM elements are
\begin{align}\label{eq:asdf}
E_\phi&=\sum_i\frac{g^2}{\pi(1+\sigma^2)}\exp\left[-g^2\frac{\modulus{\phi-\mathfrak{f}_i}^2}{1+\sigma^2}\right]\ket{e_i}\bra{e_i}.
\end{align}
We can check that the POVM elements resolve the identity
\begin{align}
\int d^2\phi E_\phi
&=  \int d^2\phi \sum_i\frac{g^2}{\pi(1+\sigma^2)}\exp\left[-g^2\frac{\modulus{\phi-\mathfrak{f}_i}^2}{1+\sigma^2}\right]\ket{e_i}\bra{e_i}
= \sum_i e^{-\frac{g^2}{1+\sigma^2}\modulus{f_i}^2}e^{\frac{g^2}{1+\sigma^2}\modulus{f_i}^2}\ket{e_i}\bra{e_i} = \Id.
\end{align}

We can rewrite \cref{eq:asdf} in terms of the efficiency $\eta$ of the noisy heterodyne measurement, this is
\begin{align}\label{eq:fdsa}
M_{\phi}
&=  \sum_i\frac{\eta g^2}{\pi}e^{-\eta g^2\modulus{\phi-f_i}^2}\ket{e_i}\bra{e_i}.
\end{align}
Both Gaussian functions in \cref{eq:asdf} and \cref{eq:fdsa}   limit to delta functions in the large gain limit.

\subsection{\texorpdfstring{$\fin$}{TEXT} is Hermitian and the linear measurement is homodyne}
We model a noisy homodyne measurement by the POVM elements
\begin{align}
M_x&=\frac{1}{\sqrt{\pi}\sigma}\int dy\,e^{-(y-x)^2/\sigma^2}\ketbra{y}{y},
\end{align}
which is the case for inefficient quadrature detection~\cite{DallDAriSacc10,Caves20_su11}. We now consider using the nonlinear amplifier unitary $V$ for Hermitian $\fin$ [generated by \cref{eq:vn_mm}] and measuring mode $a$ using the noisy homodyne POVM in \cref{eq:noisyhomapp}. Additionally mode $b$ starts in a state $\ket{\phi}_b = \int_{-\infty}^\infty dx\, \phi(x)\ket{x}_b$ with Gaussian wave function centered at $x=0$, i.e. $\phi(x) = \exp[-x^2/2\epsilon^2] /(\pi \epsilon^2)^{1/4}$.

The corresponding POVM elements are
\begin{align}
E_x = 
 \bra{\phi}_b  V^\dag M_x V  \ket{\phi}_b 
&= \bra{\phi}_b  V^\dag\left(\Id\otimes\frac{1}{\sqrt{\pi}\sigma}\int dy\,e^{-(y-x)^2/\sigma^2}\ketbra{y}{y}\right)V  \ket{\phi}_b  \\
&=\bra{\phi}_b V^\dag\left(\sum_i\ketbra{i}{i}\otimes\frac{1}{\sqrt{\pi}\sigma}\int dy\,e^{-(y-x)^2/\sigma^2}\ketbra{y}{y}\right)V\ket{\phi}_b\\
&=\sum_i\frac{1}{\sqrt{\pi}\sigma}\int dy\,e^{-(y-x)^2/\sigma^2}\bra{\phi}_b V^\dag\ket{y}\bra{y}V\ket{\phi}_b \ket{i}\bra{i}\\
&=\sum_i\frac{1}{\sqrt{\pi}\sigma}\int dy\,e^{-(y-x)^2/\sigma^2}|\ip{y-gf_i}{\phi}_b|^2 \ket{i}\bra{i}.
\end{align}
Note that $\ip{y-gf_i}{\phi}_b$ is simply the overlap between a position eigenstate and a Gaussian so $|\ip{y-gf_i}{\phi}_b|^2 = e^{-(y-gf_i)^2/\epsilon^2}/ \sqrt{\pi \epsilon^2} $. Using this we integrate the resulting Gaussian convolution
\begin{align}
E_x
&=   \sum_i \frac{1}{\sqrt{\pi}\sigma} \frac{1}{\sqrt{\pi \epsilon^2}} \int dy\,e^{-(y-x)^2/\sigma^2}e^{-(y-gf_i)^2/\epsilon^2} \ketbra{i}{i} 
=\sum_i\frac{1}{\sqrt{\pi(\sigma^2+\epsilon^2)}}\exp \left [-\frac{(x-gf_i)^2}{(\sigma^2+\epsilon^2)} \right ]\ketbra{i}{i}.
\end{align}

Changing variables to $x'=x/g$ the POVM becomes
\begin{align}
E_{x'} 
&=\sum_i\frac{g}{\sqrt{\pi(\sigma^2+\epsilon^2)}}\exp \left [-g^2\frac{(x'-f_i)^2}{\sigma^2+\epsilon^2} \right ]\ketbra{i}{i},
\end{align}
and $\sigma^2 = \frac{1-\eta }{4 \eta }$ where $\eta$ is the homodyne detection efficiency. 
This converges to a projective measurement of $f$ in the same sense as described in the previous subsection.
We can check that the POVM resolves the identity:
\begin{align}
\int dx' E_{x'}
&=\sum_i \int dx' \frac{g}{\sqrt{\pi(\sigma^2+\epsilon^2)}}\exp \left [-g^2\frac{(x'-f_i)^2}{\sigma^2+\epsilon^2} \right ]\ketbra{i}{i} 
=\sum_i \ketbra{i}{i} = \Id .
\end{align}

\section{Commutation constraints on two mode quadratic amplifiers}\label{apdx:twomodecom}
A generic quadratic operator has the form
\begin{align}
\fin&=\alpha a^2+\beta a^\dagger a+\gamma a^{\dagger 2} + \delta \Id
\end{align}
where $\alpha, \beta, \gamma,\delta \in \mathbb{C}$. To enforce normality, we require
\begin{align}
0=\comm{\fin}{\fin^\dagger}
&=\alpha\beta^*\comm{a^2}{a^\dagger a}+\alpha\alpha^*\comm{a^2}{a^{\dagger 2}}+\beta\alpha^*\comm{ a^\dagger a}{a^{\dagger 2}}+\beta\gamma^*\comm{a^\dagger a}{ a^2}+\gamma\beta^*\comm{ a^{\dagger 2}}{ a^\dagger a}+\gamma\gamma^*\comm{a^{\dagger 2}}{ a^2} \nonumber\\
&=2(\alpha\beta^*-\beta\gamma^*)a^2+2(\alpha\alpha^*-\gamma\gamma^*)\left(aa^\dagger+a^\dagger a\right)+2(\beta\alpha^*-\gamma\beta^*)a^{\dagger 2}
\end{align}
which results in the conditions $|\alpha|^2=|\gamma|^2$, $\alpha \beta^* = \beta \gamma^*$, $\beta \alpha^* = \gamma \beta^*$, and $\delta$ is unconstrained. The last two conditions are simply complex conjugates of each other, so are redundant.

\section{Statistical Efficiency of the two mode example}\label{appendix:estimators}
The heterodyne measurement is given by the POVM $E_\alpha$ and the probability of obtaining the outcome $\alpha$ is $\Pr(\alpha|\rho)$:
\begin{align}
E_\alpha=\frac{1}{\pi}\ketbra{\alpha}{\alpha}, \quad \Pr(\alpha|\rho)&=\frac{1}{\pi}\bra{\alpha}\rho\ket{\alpha}.
\end{align}
The expectation of the heterodyne outcome after amplification is:
\begin{align}
\mathbb{E}\left[(\alpha^*)^n\alpha^m\right]&=\brak{A^m(A^\dagger)^n}=\brak{(ga+\sqrt{g^2-1}b^\dagger)^m(ga^\dagger+\sqrt{g^2-1}b)^n}.
\end{align}
In particular if the $b$ mode is in vacuum:
\begin{align}
\mathbb{E}\left[\modulus{\alpha}^2\right]
=g^2\brak{a^\dagger a}+g^2 \quad {\rm and} \quad
\mathbb{E}\left[\modulus{\alpha}^4\right]
=g^4\brak{a^2(a^\dagger)^2}.
\end{align}
In the main text we defined the estimator (denoted by the hat) $\hat{n}=\frac{|\alpha|^2}{g^2}-1$, with mean $\brak{a^\dagger a}$ and variance
\begin{align}
\text{Var}\left[\hat{n}\right]=\mathbb{E}\left[\hat{n}^2\right]-\mathbb{E}\left[\hat{n}\right]^2
=\mathbb{E}\left[\frac{|\alpha|^4}{g^4}-2\frac{|\alpha|^2}{g^2}+1\right]-\brak{a^\dagger a}^2
&=\brak{a^2(a^\dag)^2}-2\brak{a^\dag a} -\brak{a^\dag a}^2 -1\\
&=\brak{(a^\dag a)^2+3a^\dag a+2}-2\brak{a^\dag a} -\brak{a^\dag a}^2 -1\\
&=\textrm{Var}\left[a^\dag a\right]+\brak{a^\dag a}  +1
\end{align}
where $\text{Var}\left[a^\dag a\right]$ is the variance of a projective measurement of the operator $a^\dag a$.

\section{An Arthurs-Kelly like model}\label{app:AK_like}
Arthurs and Kelly introduced a model for simultaneous measurements of position and momentum~\cite{ArthursKelly1965}. The model coupled the position and momentum of a single system to two independent probe systems or ``meters'' and then a subsequent measurement of those meter systems revealed information about both position and momentum. We generalize this model to provide a different measurement model for a normal operator that relies on two probe systems and subsequent position measurements. Thus the simultaneous measurements of position and momentum of a single meter in our model (heterodyne detection) gets replaced with two meters and two homodyne detectors.

\subsection{Hamiltonian and input-output relations}\label{app:AK_ham_and_io}
We have three modes now $\mathcal H_a\otimes \mathcal H_b \otimes \mathcal H_c$, the normal operator $f = (f_R+i f_I )/\sqrt{2}$ is on mode $a$ and $x_j, p_j$ for $i,j \in {b,c}$ are the position and momentum operators on $\mathcal H_j$ such that $[x_j, p_k] =i\delta_{j,k}$. The interaction Hamiltonian is
\begin{equation}
 H_{\rm m} = \kappa (f_R\otimes p_a\otimes I_c + f_I\otimes I_b\otimes p_c).
\end{equation}
Because $[p_b,p_c]=[f,p_b]= [f,p_c]=0$ and $[f_R, f_I]=0 $ the corresponding unitary $W= \exp[-i H_{\rm m} t]$ can be written as
\begin{align}
W&= \exp[-i\kappa t (f_R p_b + f_I p_c)] 
=e^{-ig f_R p_b}e^{-ig f_I p_c} = e^{-ig f_I p_c} e^{-ig f_R p_b},
\end{align}
where $g=\kappa t$ as before. The input-output relations for the meters are
\begin{align}
x_{b}^{\rm out} = W^\dag x_b^{\rm in} W = g f_R + x_b^{\rm in} 
\quad {\rm and}\quad
x_{c}^{\rm out } = W^\dag x_c^{\rm in} W = g f_I + x_c^{\rm in}, 
\end{align}
As $[f_R, x_b^{\rm in}] =0$ etc. it is easy to compute the expected noise at the output.

\subsection{Use with measurements: the effective POVM}\label{app:AK_povm}
The position wave functions $ \phi_i(x)$ of the meter states, $\ket{\phi}_i = \int_{-\infty}^\infty dx \phi(x)\ket{x}_i$, are a Gaussian centered at $x=0$ 
\begin{equation}\label{eq:gau_wf}
    \phi_i(x) = \frac{1}{(\pi \epsilon_i^2)^{1/4}} e^{-x^2/(2\epsilon_i^2)}.
\end{equation}
The ideal position (homodyne) measurement on the meters $b$ and $c$ is described by the projector  
$ P_{y_b,y_c} = \op{y_b,y_c}{y_b,y_c}$ where $x_i\ket{y_i} = y_i \ket{y_i}$. To describe inefficient detection we use the POVM $ M_{y_b,y_c} =\Id_a\otimes M_{y_b} \otimes M_{y_c}$  where $M_x$ is the noisy homodyne POVM is given in \cref{eq:noisyhomapp}.
The POVM operator for the effective measurement on $\mathcal H_a$ is
\begin{align}\label{eq:akpom}
E_{y_b,y_c} 
&= \bra{\phi_b,\phi_c} W^\dag  M_{y_b,y_c} W \ket{\phi_b,\phi_c}
= \bra{\phi_b,\phi_c} W^\dag (\Id_a\otimes M_{y_b} \otimes M_{y_c}) W \ket{\phi_b,\phi_c},
\end{align}
technically we should include the measure i.e. $dy_b dy_c E_{y_b,y_c}$. 

Our goal is for \cref{eq:akpom} to measure the normal operator $f$. First note that as $[f_R, f_I]=0$ this implies the operators can be simultaneously diagonalized
\begin{align}
f = \sum_k \mathfrak{f}_k \ket{e_k}\!\bra{e_k}
 &= \sum_k \frac{ \mathfrak{f}_k^R+i\mathfrak{f}_k^I}{\sqrt{2}} \ket{e_k}\!\bra{e_k} 
= \frac{f_R + i f_I}{\sqrt{2}} \\ 
&= \sum_k   \tilde{\mathfrak{f}}_k^R+i\tilde{\mathfrak{f}}_k^I  \ket{e_k}\!\bra{e_k} 
= \tilde{f}_R + i \tilde{f}_I \label{eq:2ndline}
\end{align}
Using this fact with the orthogonality of the eigenvectors of $f$, i.e., $\ip{e_k}{e_{k'}}=\delta_{k,k'}$, and the completeness of the eigenvectors, i.e. $\sum_k \op{e_k}{e_k}=I_a$, we find
\begin{align}
E_{y_b,y_c} 
&= \frac{1}{\sqrt{\pi}\sigma_b}\frac{1}{\sqrt{\pi}\sigma_c}\int dy_b' \int dy_c'
e^{-(y_b'-y_b)^2/\sigma_b^2}e^{-(y_c'-y_c)^2/\sigma_c^2}
|\bra{y_b'} e^{-ig f_R p_b}\ket{\phi_b}|^2 |\bra{y_c'} e^{-ig f_I p_c} \ket{\phi_c}|^2\\
&=\frac{1}{\sqrt{\pi}\sigma_b}\frac{1}{\sqrt{\pi}\sigma_c}\sum_{k,k'} \int dy_b' \int dy_c'
e^{-(y_b'-y_b)^2/\sigma_b^2}e^{-(y_c'-y_c)^2/\sigma_c^2}\times \nonumber\\
&\phantom{=\frac{1}{\sqrt{\pi}\sigma_b}\frac{1}{\sqrt{\pi}\sigma_c}\sum_{k,k'} } \Big (\op{e_k}{e_k}\otimes| \bra{y_b'} e^{-ig \mathfrak{f}_k^R p_b}\ket{\phi_b} |^2\Big) \times 
\Big(\op{e_{k'}}{e_{k'}}\otimes| \bra{y_c'} e^{-ig \mathfrak{f}_{k'}^I p_c} \ket{\phi_c}|^2 \Big )\\
&=\frac{1}{\sqrt{\pi}\sigma_b}\frac{1}{\sqrt{\pi}\sigma_c}\sum_{k} \int dy_b' \int dy_c'
e^{-(y_b'-y_b)^2/\sigma_b^2}e^{-(y_c'-y_c)^2/\sigma_c^2}
\underbrace{|\bra{y_b'} e^{-ig \mathfrak{f}_k^R p_b}\ket{\phi_b} |^2\, |\!\bra{y_c'} e^{-ig \mathfrak{f}_{k}^I p_c} \ket{\phi_c}|^2}_{:=K} \op{e_k}{e_k}.
\end{align}
Let's simplify the term labeled $K$. We use the form of the position wave function and $_i\!\bra{x}e^{-ip_i\chi} = \, _i\!\bra{x-\chi}$ we find
\begin{align}
K 
&= \int dx_b \, |\phi(x_b)|^2 |\ip{y_b'-g \mathfrak{f}_k^R}{x_b}|^2 \int dx_c \, |\phi(x_c)|^2 |\ip{y_c'-g \mathfrak{f}_{k}^I }{x_c}|^2
= \frac{e^{-(y_b'-g \mathfrak{f}_k^R)^2/\epsilon_b^2}}{\sqrt{\pi \epsilon_b^2}}  \frac{e^{-(y_c'-g \mathfrak{f}_k^I)^2/\epsilon_c^2}}{\sqrt{\pi \epsilon_c^2}}.
\end{align}
This lets us write the POVM as
\begin{align}
E_{y_b,y_c} 
&=\sum_{k}\frac{1}{\sqrt{\pi}\sigma_b}\int dy_b'
e^{-(y_b'-y_b)^2/\sigma_b^2} \frac{e^{-(y_b'-g \mathfrak{f}_k^R)^2/\epsilon^2}}{\sqrt{\pi \epsilon_b^2}}
 \frac{1}{\sqrt{\pi}\sigma_c}\int dy_c'
 e^{-(y_c'-y_c)^2/\sigma_c^2}\frac{e^{-(y_c'-g \mathfrak{f}_k^I)^2/\epsilon^2}}{\sqrt{\pi \epsilon_c^2}}
\op{e_k}{e_k}.
\end{align}
Performing both Gaussian integrals we obtain
\begin{align}
E_{y_b,y_c} 
&= \sum_{k} \frac{e^{-(y_b-g \mathfrak{f}_k^R)^2/(\epsilon_b^2+\sigma_b^2)}}{\sqrt{\pi( \epsilon_b^2+ \sigma_b^2)}}  \frac{e^{-(y_c-g \mathfrak{f}_k^I)^2/(\epsilon_c^2+\sigma_c^2)}}{\sqrt{\pi( \epsilon_c^2+ \sigma_c^2)}}   \op{e_k}{e_k}.
\end{align}
Finally we change variables (and the measure) to $\breve{y}_b=y_b/g$ and $\breve{y}_c=y_c/g$ the POVM becomes
\begin{align}
E_{\breve{y}_b,\breve{y}_c} 
&= \sum_{k} \frac{g^2 }{\pi \sqrt{( \epsilon_b^2+ \sigma_b^2)( \epsilon_c^2+ \sigma_c^2)}} \exp \left [ -g^2\frac{(\breve{y}_b-\mathfrak{f}_k^R)^2}{\epsilon_b^2+\sigma_b^2} \right ]
 \exp \left [ -g^2\frac{(\breve{y}_c-\mathfrak{f}_k^I)^2}{\epsilon_c^2+\sigma_c^2} \right ]
\op{e_k}{e_k}
\end{align}
Assuming equal squeezing ($\epsilon^2= \epsilon_b^2= \epsilon_c^2 $) and homodyne noise ($\sigma^2=\sigma_b^2=\sigma_c^2$) we obtain
\begin{align} 
E_{\breve{y}_b,\breve{y}_c} 
&= \sum_{k} \frac{g^2 }{\pi( \sigma^2 + \epsilon^2 )} \exp \left [ -g^2\frac{(\breve{y}_b-\mathfrak{f}_k^R)^2 + (\breve{y}_c-\mathfrak{f}_k^I)^2}{ \sigma^2 + \epsilon^2 } \right ]
\op{e_k}{e_k}
\end{align}
By parameterizing $\phi=\breve{y}_b +i\breve{y}_c $ and $\mathfrak{f}_k = \tilde{\mathfrak{f}}_k^R + i \tilde{\mathfrak{f}}_k^I $, i.e. the parameterization in \cref{eq:2ndline},
\begin{align}\label{eq:AKmodelapp}
E_{\phi} 
&= \sum_{k} \frac{g^2 }{\pi( \sigma^2 + \epsilon^2 )} \exp \left [ -g^2\frac{|\phi -\mathfrak{f}_k|^2 }{ \sigma^2 + \epsilon^2 } \right ]
\op{e_k}{e_k}
\end{align}
This is a valid POVM because it is manifestly positive and it is complete: $\int d\phi  E_{\phi} = \Id_a $. The complex measurement outcome $\phi$ is centered around the complex eigenvalues of $f$.

\section{Single mode example}\label{app:singlemodeexample}
In the main text we proposed nonlinear amplifier input-output relations of the form
$\aout = \gamma \fin(x_\theta) + \tilde \mu a_\theta + \tilde\nu a_\theta\dg$.
Using a parametrization similar to Ref.~\cite{SienLisi01,*WuCote02,*DellSiena04,*BrunHouh19} we have
\begin{align*}
\gamma &=g e^{i\varphi} & a_\theta &= a e^{i\theta}\\
\mu &= \cosh r & \tilde \mu &= \mu e^{i\theta}\\
\nu &= -e^{2i \phi}\sinh r& \tilde \nu &= \nu e^{i\theta}.
\end{align*}
The commutation relations to be preserved from input to output when ${\rm Re}[\tilde \mu \gamma^* - \tilde \nu^* \gamma]=0$. 
This becomes
\begin{align*}
0={\rm Re}[\tilde \mu \gamma^* - \tilde \nu^* \gamma]
= g \cosh r \cos (\theta -\varphi )+ g \sinh r \cos (\theta -\varphi -2 \phi )
= \cosh r \cos (\theta -\varphi )+  \sinh r \cos (\theta -\varphi -2 \phi ),
\end{align*}
note the sign here is different from previous works~\cite{SienLisi01,*WuCote02,*DellSiena04,*BrunHouh19} because we have a different convention for the sign of the squeeze coefficients, specifically $\nu$. 
Nevertheless the condition is satisfied if 
$\cos (\theta -\varphi )=0$ and $\cos (\theta -\varphi -2 \phi )=0$.
Now lets set $\theta =0$ so that
$\cos (-\varphi )=\cos (\varphi )=0$ and $\cos (-\varphi -2 \phi ) = \cos (\varphi +2 \phi ) =0$.
Thus $\varphi = \pm \pi/2$ is a valid solution for the first equation. We choose the $\varphi =  \pi /2$ solution. This means
$\cos (-\pi/2 +2 \phi ) = -\sin (2 \phi )=0$
which in turn means we can take $\phi=0$. Summarizing we have
\begin{align}
\theta = 0 \quad \text{(rotated\ quadrature\ angle)},\quad 
\varphi =  \pi /2 \quad \text{(gain\ phase)}, \quad 
\phi = - \pi /2 \quad \text{(squeezing\ angle)}.
\end{align}
Using these parameters we obtain \cref{eq:specialcase_single_mode} in the main text: $\aout = igf(x)+\cosh r a+\sinh r a^\dag$.
It is easy to verify that  $\aout$ preserves the commutation relations 
\begin{align}
\comm{\aout}{\aout^\dag} 
&=ig\left((\sinh r+\cosh r)\comm{f(x)}{a}+(\cosh r+\sinh r)\comm{f(x)}{a^\dag}\right)+1\\
&=ig(\sinh r+\cosh r)\left(\comm{f(x)}{a}+\comm{f(x)}{a^\dag}\right)+1\\
&=\sqrt{2}ig(\sinh r+\cosh r)\comm{f(x)}{x}+1\\
&= 1
\end{align}

The output quadrature 
\begin{align}
\xout &= \frac{(\aout+\aout\dg)}{\sqrt{2}} 
=  ( \cosh r \ain + \sinh r \ain\dg+\cosh r \ain\dg + \sinh r \ain)
=   e^{r}( \ain + \ain\dg )/ \sqrt{2}
=   e^{r} \xin.
\end{align}
The orthogonal output quadrature, i.e. $\pout = -i(\aout-\aout\dg)/\sqrt{2}$, is 
\begin{align}
\pout 
&= \frac{-i}{ \sqrt{2}} \left (i {g}f(\xin) + \cosh r \ain + \sinh r \ain\dg
+i {g}f(\xin) - \cosh r \ain\dg - \sinh r \ain\dg \right) 
=  \sqrt{2}g f(\xin)+  e^{-r} \pin.
\end{align}

Now we work out the symmetrically ordered noise
\begin{align}
\expt{|\Delta \pout |^2} 
&=  2g^2 \expt{f(\xin)^2} + \sqrt{2}g e^{-r}\expt{f(\xin) \pin + \pin f(\xin)}+ e^{-2r} \expt{\pin^2} \nn\\
&\phantom{=}-\left [ 2g^2 \expt{f(\xin)}^2 + \sqrt{2}g  e^{-r}\left ( \expt{f(\xin)}\expt{\pin}+ \expt{\pin}\expt{f(\xin)}\right ) +  e^{-2r} \expt{\pin}^2
\right ]\\
&=  2g^2 \expt{|\Delta f(\xin)|^2} + 
 \sqrt{2}g e^{-r} \left ( \expt{f(\xin)\pin + \pin f(\xin)} -  ( \expt{f(\xin)}\expt{\pin}+ \expt{\pin}\expt{f(\xin)} )     \right )+ 
e^{-2r} \expt{|\Delta \pin|^2}
\end{align}
In the large squeezing limit we obtain the results stated in the main text.

\end{document}